\pdfoutput=1

\documentclass[journal,twoside]{IEEEtran}

\IEEEoverridecommandlockouts                              

\usepackage{flushend}  
\usepackage{ifpdf}
\usepackage{cite}
\ifCLASSINFOpdf
  \usepackage[pdftex]{graphicx}
  \DeclareGraphicsExtensions{.pdf,.jpeg,.png}
\else
  \usepackage[dvips]{graphicx}
  \DeclareGraphicsExtensions{.eps}
\fi
\graphicspath{{./figures/}}
\usepackage[cmex10]{amsmath}
\usepackage{amsfonts}  
\usepackage{amssymb}  
\usepackage{amsthm}
\usepackage{array}
\usepackage{algorithm}
\usepackage{algpseudocode}

\newcommand{\vect}[1]{\mathbf{#1}}
\newcommand{\matr}[1]{\mathbf{#1}}

\DeclareMathOperator{\sign}{sign}

\theoremstyle{plain}

\theoremstyle{definition}

\theoremstyle{remark}

\makeatletter
\newlength \figwidth
\if@twocolumn
\else
\fi
\makeatother

\usepackage{hyperref}
\hypersetup{
  pdfinfo={
    Title={Numerical Issues Affecting LDPC Error Floors},
    Author={Brian K. Butler},
    Subject={Error Correction Coding; Signal Processing},
    Keywords={LDPC codes; Error Correction; Error Floor; Message Passing Decoding}
  }
}

\begin{document}
\title{Numerical Issues Affecting LDPC Error Floors}
\author{
\IEEEauthorblockN{Brian~K.~Butler and Paul~H.~Siegel}\\
\IEEEauthorblockA{Department of Electrical and Computer Engineering, University of California San~Diego, La~Jolla, CA 92093\\
butler@ieee.org, psiegel@ucsd.edu}%
\thanks{This research was supported in part by NSF Grants CCF-0829865 and CCF-1116739 and by the Center for Magnetic Recording Research at UCSD.}
}

\maketitle

\ifCLASSOPTIONpeerreview
	\markboth{Numerical Issues Affecting LDPC Error Floors}%
	{Numerical Issues Affecting LDPC Error Floors}
\else
	\markboth{\MakeLowercase{submitted to} \textit{IEEE GLOBECOM 2012} \;\;\;\;Version:  AUGUST 1, 2012}%
	{Butler and Siegel: Numerical Issues Affecting LDPC Error Floors}
\fi

\begin{abstract} 
Numerical issues related to the occurrence of error floors in floating-point simulations of belief propagation (BP) decoders are examined. Careful processing of messages corresponding to highly-certain bit values can sometimes reduce error floors by several orders of magnitude. Computational solutions for properly handling such messages are provided for the sum-product algorithm (SPA) and several variants. 

\end{abstract}


%
\IEEEpeerreviewmaketitle

\section{Introduction} 
Belief propagation (BP) decoders based upon the sum-product algorithm (SPA) are widely used to decode error-correcting codes that have a sparse graphical representation, including, most notably, low-density parity-check (LDPC) codes.
This paper addresses numerical issues arising in the implementation of SPA decoding and several of its variants that have an impact on the occurrence and severity of frequently observed error floors in the decoder performance curves.
(For  background on BP and LDPC coding, the reader is referred to \cite{Gal63,RichUrb,RyanLin}.)

In \cite{MKfloor}, MacKay and Postol examined  \emph{near codewords} associated with the error 
floors that they observed in BP decoding of Margulis-type codes over the additive white Gaussian noise (AWGN) channel. Shortly thereafter, Richardson \cite{RichFloors} wrote a seminal paper on the error floors of LDPC codes in the setting of the AWGN channel and the binary symmetric channel, identifying decoder-dependent, error-prone substructures in the Tanner graph, dubbed \emph{trapping sets}, that were responsible for the observed floors.  Later, Han and Ryan also studied error floors and their properties for LDPC codes used on the AWGN channel \cite{Han09}. With the exception of \cite{RichFloors}, which used a 5-bit hardware simulation to generate error-rate curves, it appears that the error floors reported in these prior studies were found with floating-point (FP) simulations of the BP decoder. 

A quantization scheme, such as FP, imposes a limited range and finite resolution on the values to be represented.
Additionally, the use of non-linear functions in a quantized environment may dramatically further limit 
the domains and/or images of said functions.  
This paper addresses these issues of range and resolution of the messages in several SPA variants implemented in FP.
Recent work has shown that the error floors of variable-regular LDPC codes, such as the Margulis code considered by the authors of \cite{RyanLin,MKfloor,RichFloors,Han09}, are largely the result of numerical problems associated with the processing of highly certain messages in the BP decoder implementation~\cite{SunPhD,ButlerAller,ButlerFloor}.  

This work improves the range of several known SPA variants, introduces a new SPA variant,
and presents simulation results for two variable-regular LDPC codes whose error floors are reduced by orders of magnitude by addressing these numerical range issues.
Motivated by these observations, Zhang and Siegel~\cite{XiaojieFloor} have devised several small-bit-width decoders that
substantially lower the observed error floors of several LDPC codes.
These decoders use non-uniform quantization techniques 
that accept reduced message resolution for highly certain messages for the sake of increased message range.

\section{Background} 
\subsection{Code and Channel} 
LDPC codes are defined by the null space of a parity check matrix $\matr{H}$.
The codewords are the set of column vectors $\mathcal{C}$, such that $\vect{c} \in \mathcal{C}$ satisfies 
$\matr{H} \vect{c} = \vect{0}$ over a particular field.
A given code can be described by many different $\matr{H}$ matrices.

The $\matr{H}$ matrix over GF($2$) may be associated with a bipartite graph $B=(V,C,E)$, called a \emph{Tanner graph}, in which the vertex set may be partitioned into two disjoint sets $V$ and $C$.
The set of variable nodes $V$ represent the symbols of the codeword that are sent over the channel and correspond to the columns of the parity-check matrix.
The set of check nodes $C$ enforce the parity-check equations represented by the rows of $\matr{H}$.
Each edge $e\in E$ of a Tanner graph joins a variable node $v_i \in V$ to a check node $c_j \in C$. 
The $(j,i)$th entry of $\matr{H}$ is $1$ if $(v_i,c_j) \in E$ and is $0$ otherwise.

\theoremstyle{definition}
\newtheorem*{assn}{Assumption} 
\begin{assn}
We are only concerned with codes over the binary field GF($2$).
Also, we assume binary antipodal signaling over the AWGN channel for the purposes of illustration, but
the fundamental numerical issues are independent of these assumptions.
\end{assn}

After encoding, each binary element of codeword $\vect{c}$ is transmitted over the AWGN channel
as a binary antipodal symbol $t_i \in \{+1,-1\}$. Every received symbol $r_i$ is simply
the sum $t_i+n_i$ of the transmitted symbol plus independent and identically-distributed (i.i.d.) Gaussian noise
of zero-mean and variance $\sigma^2$. 
This yields a channel SNR of $1/\sigma^2$ or $2 R E_b / N_0$, where $R$ is the rate of the code,
$E_b$ is the received energy per information bit for coherent bandpass detection, and
$N_0$ is the one-sided power spectral density of the noise.

\subsection{Sum-Product Algorithm (SPA)} 
We next describe the sum-product algorithm (SPA) that is one of the most widely used forms of BP decoding for LDPC codes.  
For codes with a cycle-free Tanner graph representation, SPA decoding on such a graph is optimal, in the sense that it is equivalent to symbol-wise, maximum a-posteriori (MAP) decoding. 
Such codes are generally not attractive \cite{cyclefree}.
Although no longer optimal when used to decode LDPC codes whose Tanner graph representations have cycles, SPA decoding has nevertheless been found to provide near-optimal performance, at least in certain ranges of signal-to-noise ratio.

As is often done, we will first implement our decoder simulation in the log-domain; this removes the need for normalization and is also closer to the approximations used in hardware implementations.
In describing the algorithm, we use the notation $\mathcal{N}(i)$ to indicate the set of check nodes adjacent to variable node $v_i$, and we use $\mathcal{N}(j)\setminus i$ to denote all variable nodes adjacent to check node $c_j$, excluding variable node $v_i$.

First, the received symbols $r_i$ are converted to log-likelihood ratios (LLRs)
defined by
\begin{equation*}
\lambda^{[i]} = \ln \frac{{P({r_i}|{t_i} = + 1)}}{{P({r_i}|{t_i} = - 1)}}.
\end{equation*}
For the AWGN channel, this becomes $\lambda^{[i]} = 2 r_i / {\sigma ^2}$.

The LLR $\lambda^{[i]}$ is the initial message passed from variable node $v_i$
to each of its adjacent check nodes in the Tanner graph. That is, $\lambda_{0}^{[i \to j]} = \lambda^{[i]}$  for all $i, j$ such that $(v_i,c_j) \in E$. The iteration counter $l$ is then initialized to $1$.

On the first half-iteration of iteration $l$, the message sent from check node $c_j$ to the adjacent variable node $v_i$ is given by
\begin{equation}
\label{CNupdate}
\lambda_{l}^{[i \leftarrow j]} = 2\, \tanh^{-1} \Biggl[ {\prod\limits_{k \in \mathcal{N}(j)\setminus i} {\tanh \frac{{\lambda_{l - 1}^{[k \to j]}}}{2}} } \Biggr],
\end{equation}
for each edge in the graph.
During the second half-iteration, the return message sent from  variable node $v_i$ to adjacent check node $c_j$ is given by 
\begin{equation}
\label{VNupdate}
\lambda_{l}^{[i \to j]} = \lambda^{[i]} + \sum\limits_{k \in \mathcal{N}(i)\setminus j} {\lambda_{l}^{[i \leftarrow k]}}.
\end{equation}
If the early termination logic detects a codeword or if the iteration counter $l$ exceeds the maximum allowed count, the iterations are halted. 
Otherwise, the iteration counter is incremented by 1, and the new check-to-variable-node messages are computed using (\ref{CNupdate}).
Upon the completion of the iterations, the decision for symbol $i$, denoted $\hat{t}_l^{[i]}$, is set equal to the sign of the incoming message sum at variable node $v_i$, that is, 
\begin{equation*}
\hat{t}_l^{[i]} = \sign\Bigl(\lambda^{[i]} + \sum\limits_{k \in \mathcal{N}(i)} {\lambda_{l}^{[i \leftarrow k]}}\Bigr).
\end{equation*}
If $B$ has no cycles, the message sum is equivalent to $\ln \frac{{P( \vect{r}|{t_i} = + 1)}} {{P( \vect{r}|{t_i} = -1 )}}$, the MAP decision statistic.


\subsection{Floating-Point Notation} 
Floating-point (FP) formatting offers an economical computer representation (\textit{i.e.}, quantization) of real numbers covering 
a wide dynamic range with a significant level of precision.

The elements of an FP number are stored separately: the sign bit, the exponent, and the significand.
Table \ref{table_flt} shows the basic binary formats included in IEEE Standard 754-2008, which uses base-$2$ representation \cite{754}.
The parameters $p$ and $emax$ denote the number of binary digits in the significand ({i.e.}, \emph{precision}) and the maximum exponent, respectively.
The standard requires the minimum exponent be $emin = 1 - emax$.
Since normalized FP numbers are expressed with the radix point after the first binary digit of the significand, 
the maximum supported FP value may be computed to be
\begin{equation*}
\overbrace{(1.111\ldots111)_2}^{p \text{ ones}} \cdot 2^{emax}= (2-2^{1-p})\ 2^{emax}.
\end{equation*}

\begin{table}[!t]
\renewcommand{\arraystretch}{1.15} 
\caption{Floating-point basic binary formats of IEEE Standard 754 \cite{754}.}
\label{table_flt}
\centering
\begin{tabular}{l||r||r||r}
\hline
\bfseries Name & \bfseries Bits stored & $p$ \bfseries bits & $emax$\\
\hline\hline
Single-precision      	(SP)& $32$ & $24$ & +127\\
Double-precision   	(DP)& $64$ & $53$ & +1023\\
Quadruple-precision	 (QP)& $128$ & $113$ & +16,383\\
\hline
\end{tabular}
\end{table}

Normalization of binary FP numbers is required by the standard when possible.
These \emph{normalized} numbers must have $1$ as the most significant bit of the significand, which need not be stored.
Very small FP numbers may become subnormal (or ``denormalized'') when they are too small to be normalized.
Subnormal numbers are supported by the standard and do not use the full precision available. 
The smallest normalized positive value is $2^{emin}$ and the smallest subnormal positive value is
\begin{equation*}
\overbrace{(.000\ldots001)_2}^{p-1 \text{ bits}} \cdot 2^{emin}= 2^{emin+1-p}
\end{equation*}
per \cite[\textsection 3.3]{754}.
The available binary FP values in the immediate vicinity of $1$ are the following:
\begin{equation}
\label{nearone}
1-3\cdot2^{-p}, 1-2\cdot2^{-p}, 1-2^{-p}, 1, \text{ and } 1+2^{1-p}.
\end{equation}



\subsection{Numerical Problem in FP} 
\label{sectNP}
Direct implementation of (\ref{CNupdate}) yields numerical problems at high LLRs.
Letting $y=\tanh(\lambda/2)<1$, we may express $y$ as
\begin{equation*}
y=\frac{1-e^{-\lambda}}{1+e^{-\lambda}}=1 - \frac{2e^{-\lambda}}{1+e^{-\lambda}}
\end{equation*}
in order to find when FP quantizes $y$ to $1$. 
Ideally, the values of $y$ near $1$ are rounded to the closest quantized level listed in (\ref{nearone}).
So to be rounded-up to $1$, $y\ge1-2^{-p}/2$ must be satisfied, which is equivalent to
\begin{align}
\label{SPAprob2}
\frac{2e^{-\lambda}}{1+e^{-\lambda}} &\le 2^{-(p+1)}\text{ and}\\
\label{SPAprob3}
\lambda + \ln(1+e^{-\lambda}) &\ge (p+2) \ln{2}.
\end{align}
We may very accurately approximate (\ref{SPAprob3}) by $\lambda\ge(p+2)\ln2$ or $38.1230949$
in DP-FP ($64$-bit IEEE 754) with its $p=53$ bits of precision.

As an argument of $\pm 1$ will cause the $\tanh^{-1}$ function to overflow, protection from high magnitude
LLRs must be added to (\ref{CNupdate}) or (\ref{VNupdate}) to ensure numerical integrity or an alternative
solution not using $\tanh^{-1}$ must be found.
Thus, preventing $\tanh^{-1}$ overflow by limiting LLRs will result in a maximum producible LLR magnitude.
Our examination of published error-floor results suggests that such LLR limiting (or ``saturating'' or ``clipping'') is commonly employed.

\section{Preferred SPA Solution}
\label{sect-SPAwo}
In this section we describe our preferred SPA solution, its numerical limits, and speed issues.
The relationship known as the Jacobian logarithm is 
\begin{equation}
\label{eqsplot}
\begin{split}
\ln(e^x+e^y) &= \ln(e^x) + \ln\left(1+e^{y-x} \right)\\
	&= \ln(e^y) + \ln\left(1+e^{x-y} \right)\\
	&= \max(x,y) + \ln\left(1+\exp( - |x-y| ) \right).
\end{split}
\end{equation}
Using (\ref{eqsplot}) and other identities, an alternative exact pairwise check node reduction may be derived \cite{HuSPA,AnastSPA}.
For example, if we set $\mathcal{N}(j)\setminus i = \{k_1,k_2\}$, the check-node message in (\ref{CNupdate}) can be computed as follows:
\begin{equation}
\begin{split}
\label{CNupdate2}
\lambda_{l}^{[i \leftarrow j]} =
 &\sign{\lambda_{l - 1}^{[k_1 \to j]}} \cdot
 \sign{\lambda_{l - 1}^{[k_2 \to j]}} \cdot \\
 &\min\left( \left| \lambda_{l - 1}^{[k_1 \to j]} \right|, \left| \lambda_{l - 1}^{[k_2 \to j]} \right| \right)\\
 &+\ln \left(1+ \exp \left[ - \left| \lambda_{l - 1}^{[k_1 \to j]}+\lambda_{l - 1}^{[k_2 \to j]} \right| \right] \right)\\
 &-\ln \left(1+ \exp \left[ - \left| \lambda_{l - 1}^{[k_1 \to j]}-\lambda_{l - 1}^{[k_2 \to j]} \right| \right] \right)
\end{split}
\end{equation}
The check-node-message re-formulation in (\ref{CNupdate2}) contains no possibility of overflow, regardless of the LLR magnitude, which for DP-FP extends to approximately $1.798 \times 10^{308}$.
The only potential for overflow in the SPA is now just the addition operation within (\ref{VNupdate})
at extremely high LLRs.

The computer implementation of (\ref{CNupdate2}) does not necessarily have a large impact on simulation speed.
A single hyperbolic tangent evaluation consumes nearly the same number of CPU cycles
as four exponential or logarithmic evaluations on a modern processor in our experiments.
The most significant impact is that (\ref{CNupdate2}) forces us to organize computations pairwise.
Hu {et al.} present a substantial speed improvement by organizing computation pairs onto
a trellis by the forward-backward algorithm, which computes \emph{all} the output messages of the check node \cite{HuSPA}.

The next level of potential speed optimization is to approximate the $\ln\left(1+\exp(-|x|)\right)$ operations,
for which results lie in the interval $(0, \ln 2]$.
There has been significant work in this area, much of it focused towards hardware implementation \cite{HuSPA, ChenFoss}.
For simulation, we have often adopted the following two-piece linear approximation of Richter {et al.} \cite{Richter05}.
\begin{equation*}
\ln\left(1+ e^ {-\left| x \right|}\right) \approx \begin{cases}
0.6 - 0.24\left| x \right|, &\mbox{if $\left| x \right| <2.5$}\\
0,                                &\mbox{otherwise}.
\end{cases}
\end{equation*}
We have noted losses of about $0.02$ dB in the AWGN waterfall region of the FER curve for the $(2640,1320)$ Margulis code with this approximation,
while the simulation runs nearly $4$ times faster.
This is an acceptable trade-off for a decoder simulation tool used to study the error floor.

\section{Additional SPA Formulations} 
In this section we address the numerical issues of other versions of the SPA implemented using FP computer processing. 
As explained in Section \ref{sectNP}, the $\tanh(\lambda/2)$ function loses accuracy and rounds to $1$ for $\lambda\ge(p+2)\ln2$.
Thus, changing to larger FP formats for added precision increases the LLR limits only linearly.
This section explores alternative formulations of SPA, some of which increase LLR dynamic range.

\subsection{Min-Sum Algorithm (MSA)}
The min-sum algorithm (MSA) uses the following approximation to the check-node-update expression (\ref{CNupdate}):
\begin{equation}
\label{CNupdateMSA}
\lambda_{l}^{[i \leftarrow j]} = \min_{k \in \mathcal{N}(j)\setminus i}{ \left| \lambda_{l - 1}^{[k \to j]} \right|} \cdot
\prod\limits_{k \in \mathcal{N}(j)\setminus i} \sign \lambda_{l - 1}^{[k \to j]}.
\end{equation}
Since this expression is equivalent to (\ref{CNupdate2}) with the two logarithmic terms assumed to be zero, 
(\ref{CNupdateMSA}) also won't overflow.
However, the decoder performance losses that have been observed by using MSA have ranged from 0.60 to 1.22 dB \cite{ChenFoss} in AWGN, 
depending on the code.
To reduce these losses, variations on (\ref{CNupdateMSA}) known as normalized-BP and offset-BP have been used successfully \cite{RyanLin,ChenFoss}.
Note that as LLRs get very large, the MSA closely approximates the SPA as the two logarithmic terms of (\ref{CNupdate2}) become relatively small.

\subsection{Gallager's Involution Transform (GIT)}
In \cite{Gal63}, Gallager proposed another version of the check-node-update expression (\ref{CNupdate}), using logarithms to replace the
multiplications with additions. 
The resulting check-node update becomes
\begin{equation*}
\begin{split}
\lambda_{l}^{[i \leftarrow j]} = \prod\limits_{k \in {\cal N}(j)\setminus i} & \sign \left( {\lambda_{l-1}^{[k \to j]}} \right) \cdot\\
&\Phi \biggl( {\sum\limits_{k \in {\cal N}(j)\setminus i} {\Phi \left( {\left| {\lambda_{l-1}^{[k \to j]}} \right|} \right)} } \biggr),
\end{split}
\end{equation*}
where we define
\begin{equation}
\label{Invol2}
\Phi \left( x \right) \triangleq - \ln \tanh \left( \frac{x}{2} \right) = \ln \left( {\frac{{1 + {e^{ - x}}}}{{1 - {e^{ - x}}}}} \right),
\end{equation}
for $x>0$. 
Since the function $\Phi \left( x \right)$ is its own inverse, \textit{i.e.}, $\Phi \left( \Phi \left( x \right) \right) = x$, for all $x>0$,
this technique is sometimes called Gallager's involution transform (GIT).
Because the function $\Phi \left( x \right)$  transforms values between domains in which addition 
is the primary means of computing,   
this technique was originally proposed for low-complexity hardware implementation of the SPA.
We are also aware of its appearance in SPA simulation code, 
in spite of the fact that addition is no faster than multiplication on a modern FP processor.

Both expressions in (\ref{Invol2}) suffer finite-precision problems prior to taking the logarithm.
As explained in Section \ref{sectNP}, $\tanh \left( \lambda/2 \right)$ is rounded to $1$ for $\lambda\ge(p+2) \ln2$, 
which would yield $\Phi \left( \lambda \right)=-\ln1=0$.
Similarly, $1 - e^{-x}$ and $1 + e^{-x}$ are rounded to $1$ for $x\ge(p+1) \ln2$ and $x>p \ln2$, respectively.
Thus, the computational limit of LLR magnitude using (\ref{Invol2}) is at best $\lambda\ge(p+2) \ln2$.

\subsection{Gallager's Involution Transform, Amended (GIT2)}
\label{sectGIT2}
Given the following series expansion in the range $x>0$:
\begin{equation*}
\ln \left( {1 + {e^{ - x}}} \right) = e^{-x} - \frac{e^{-2x}}{2} + \frac{e^{-3x}}{3} + \ldots,
\end{equation*}
the series expansion of (\ref{Invol2}) is readily found to be
\begin{equation}
\label{Invol3}
\ln \left( {\frac{{1 + {e^{ - x}}}}{{1 - {e^{ - x}}}}} \right) = 2 \left[ e^{-x} + \frac{e^{-3x}}{3} + \frac{e^{-5x}}{5} + \ldots \right],
\end{equation}
for $x>0$.
This can clearly be approximated by a small number of terms as $x$ grows large.
What has been gained is that (\ref{Invol3}) will not round to zero until $x>(emax+p)\ln2$, thus providing a substantial increase in LLR dynamic range.
Also, the value $e^{-x+\ln2}$ will not begin to lose accuracy as it ``denormalizes'' until $x > emax \ln2$. 

Our next step is to find the cross-over in accuracy between (\ref{Invol2}) and (\ref{Invol3}) with a limited number of terms.
Make the following computations:
\begin{alignat*}{2}
\Phi_0(x) &\triangleq - \ln \tanh \left( \frac{x}{2} \right),\\
\Phi_1(x) &\triangleq 2 e^{-x},\\
\Phi_2(x) &\triangleq 2e^{-x} + \frac{2e^{-3x}}{3},\quad\mbox{and}\\
\epsilon_i(x) &\triangleq \left| \Phi_i(x) - \Phi(x) \right| / \Phi(x),
\end{alignat*}
where we compute $\Phi (x)$ using our best approximation at each $x$ and not (\ref{Invol2}) explicitly.
The expression for $\epsilon_i(x)$ computes the relative error of the corresponding computation $\Phi_i(x)$.

\begin{figure}
\centering
\includegraphics[width=\figwidth]{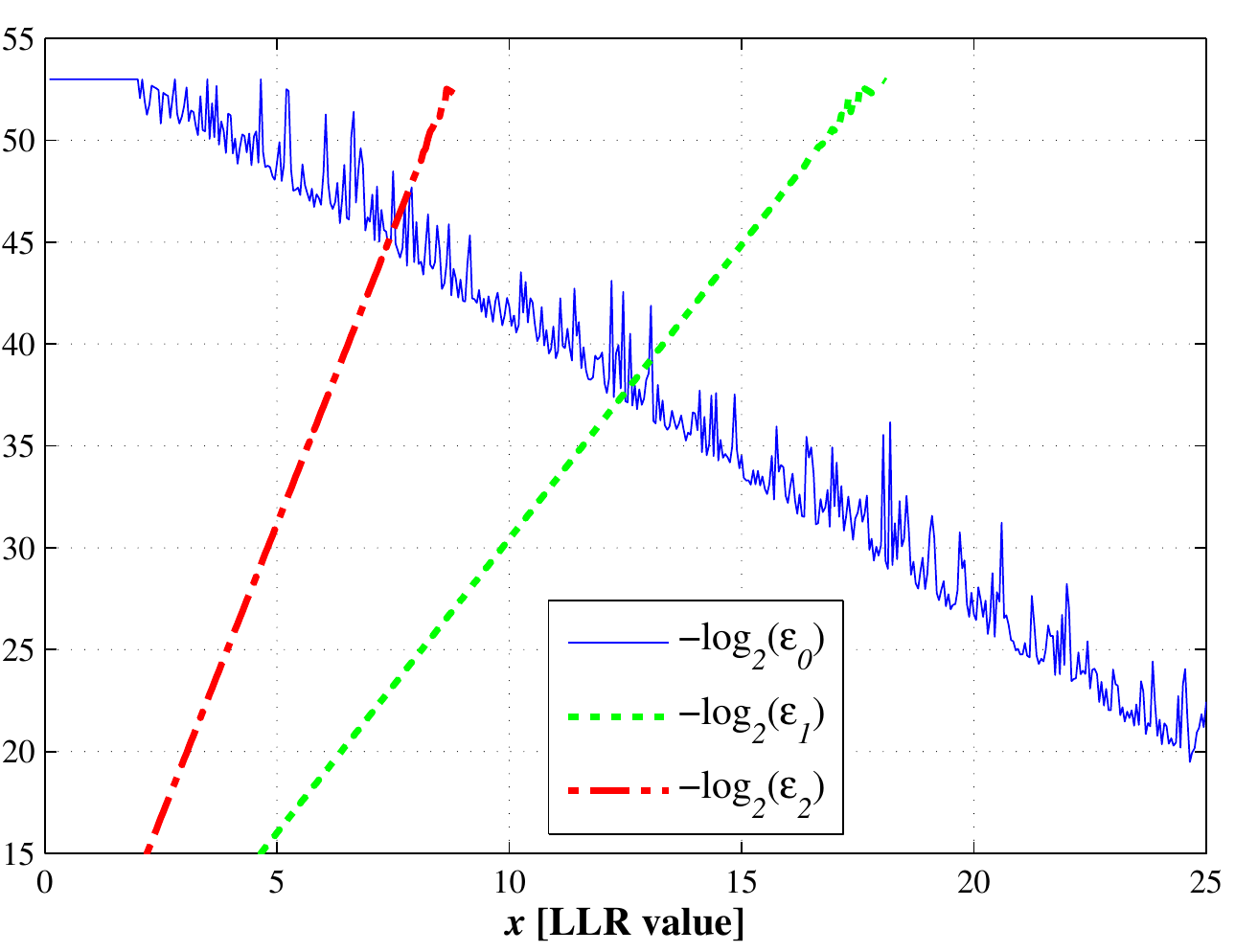}
\caption{Numerical accuracy of several versions of $-\ln\tanh(x/2)$ plotted as significant bits vs. $x$ (LLR value) using DP-FP computations.}
\label{fig_lntanh}
\end{figure}

In Fig.\ \ref{fig_lntanh} we plot the relative error of each computation as $-\log_2(\epsilon_i)$ versus $x$ to show the accuracy in bits 
of the several computational versions of $\Phi(x)$ in DP-FP.
We observe a cross-over of the lower bound of the accuracy of $\Phi_0(x)$ and the accuracy of the single-term power series $\Phi_1(x)$ 
at $x \approx 12.4$ and $37.2$ bits of accuracy, which is still quite good.
Thus, computationally, the following approximation to (\ref{Invol2}) has much greater dynamic range 
than directly implementing (\ref{Invol2}) on a computer, at acceptable accuracy for many applications:
\begin{equation}
\label{Invol5}
\Phi (x) \approx \begin{cases}
 -\ln\tanh \left( \frac{x}{2} \right),  &0<x<12.4\\
                              e^{-x+\ln2},   &x\ge12.4.
\end{cases}
\end{equation}
The value returned by (\ref{Invol5}) will not round to zero until $x>(emax+p)\ln 2$ or $745.8$ in DP-FP.
If even more accuracy is desired, additional terms may be employed at an earlier cross-over.
For instance, Fig.\ \ref{fig_lntanh} shows that the cross-over in the accuracy of $\Phi_0(x)$ and the two-term power series $\Phi_2(x)$ occurs at $x \approx 7.35$ 
with over $44$ bits of accuracy.
However, no larger dynamic range is achievable while employing Gallager's involution transform.

Our proposed computation (\ref{Invol5}) achieves a factor of $(emax+p)/(p+2)$ increase in the LLR dynamic range of (\ref{Invol2}), 
which is approximately a $20$-fold improvement for DP-FP.
Also, at LLR magnitudes larger than $12.4$, this simple modification achieves higher accuracy with reduced computational complexity than (\ref{Invol2}).

\subsection{Likelihood Ratio (LR)}
\label{sectLR}
An alternative to using $\tanh$ in the SPA is to perform the required computations in the likelihood ratio (LR) domain \cite{Kschischang,And_shortcomings}.
Interestingly, the non-uniform FP quantization of LRs maps to nearly uniform quantization in the LLR domain.
If we let $L^{[i]} \triangleq \exp{\lambda^{[i]}}$,
$L_l^{[i\to j]} \triangleq \exp{\lambda_l^{[i\to j]}}$, and so on,
the variable-node update becomes
\begin{equation}
\label{LR_VNupdate}
L_{l}^{[i \to j]} = L^{[i]} \prod\limits_{k \in \mathcal{N}(i)\setminus j} {L_{l}^{[i \leftarrow k]}} \quad\text{or}
\end{equation}
\begin{equation}
\label{LR_VNupdate2}
L_{l}^{[i \to j]} = \exp\Bigl( \ln L^{[i]} + \sum\limits_{k \in \mathcal{N}(i)\setminus j} {\ln L_{l}^{[i \leftarrow k]}} \Bigr).
\end{equation}
Letting $\mathcal{N}(j)\backslash i = \{k_1,k_2\}$, 
the pairwise check-node update may be computed as
\begin{equation}
\label{CNupdate4}
L_{l}^{[i \leftarrow j]} 
= \frac {1 + L_{l - 1}^{[k_1 \to j]} L_{l - 1}^{[k_2 \to j]}}{L_{l - 1}^{[k_1 \to j]} + L_{l - 1}^{[k_2 \to j]}}.
\end{equation}
Andrews \cite{And_shortcomings} notes that multiplicative overflow within (\ref{CNupdate4}) is avoided
if the input LRs are limited to $L_{l - 1}^{[k \to j]} < \sqrt{2^{emax} (2-2^{1-p})}$, 
which corresponds to an LLR limit of about $354.89$ in DP-FP.
We have found a numerical improvement to this computation which doubles its LLR range.
That improvement is to appear in the full version of this paper.

Note that the variable-node update (\ref{LR_VNupdate}) is more sensitive to multiplication overflow than (\ref{CNupdate4}) for $d_v \ge 3$.
However, (\ref{LR_VNupdate}) may be re-stated as (\ref{LR_VNupdate2}) which has no intermediate overflows.
We may gracefully limit the argument of the exponential in (\ref{LR_VNupdate2}) to restrict the output LR range and prevent overflow.
Of course, (\ref{LR_VNupdate2}) is more complex to evaluate.



\subsection{Likelihood Difference (LD) or Tanh Domain}
\label{sectLD}
Another alternative is to perform the computations in the likelihood difference (LD) domain, on the interval $(-1,+1)$ \cite{Kschischang}. 
This is variously known as the $\tanh$ or soft-bit domain \cite{tanhdomain}.
If we let $\delta^{[i]} \triangleq {{P({r_i}|{t_i} = - 1)}}-{{P({r_i}|{t_i} = + 1)}}$ and so on, 
the pairwise variable-node update becomes
\begin{equation}
\label{LD_VNupdate}
\delta_{l}^{[i \to j]} = \frac{\delta_{l}^{[i \leftarrow k_1]} + \delta_{l}^{[i \leftarrow k_2]}}
        {1+\delta_{l}^{[i \leftarrow k_1]} \delta_{l}^{[i \leftarrow k_2]}},
\end{equation}
while the check-node update is simply
\begin{equation}
\label{LD_CNupdate}
\delta_{l}^{[i \leftarrow j]} = \prod\limits_{k \in \mathcal{N}(j)\setminus i} \delta_{l-1}^{[k \to j]}.
\end{equation}
The dominant numerical issue for the check-node update is the resolution of $\delta$ in the neighborhood of $\delta=\pm1$.  
The LD messages closest to absolute certainty in FP are $\pm(1-2^{-p})$.
Since LLRs are related to LDs by $\lambda = \ln(1+\delta) - \ln(1-\delta)$, the greatest nearly certain message available in LD
is equivalent to an LLR magnitude of 
\begin{equation}
\begin{split}
\label{LDlim}
\lambda &= \ln(2-2^{-p}) - \ln(2^{-p}) \\
&= (p+1) \ln 2 + \ln(1-2^{-p-1}) \approx  (p+1) \ln 2,
\end{split}
\end{equation}
where our approximation error is less than the available resolution of FP.
Note that the variable-node update (\ref{LD_VNupdate}) may suffer from rounding to $\pm1$ issues and divide by $0$ errors for highly certain messages,
but our focus in this section has been on the limitations of the check-node updates.
Thus, for comparative purposes, we use (\ref{LDlim}) to represent the equivalent LLR limits of this formulation.


\subsection{Offset-Likelihood Difference (OLD)}
\label{sectOLD}
The magnitude of a quantization error in (normalized) FP is roughly proportional to the amplitude of the represented value.
Thus, FP quantizes with greater absolute accuracy close to $0$ than close to $1$.
This leads us to propose that LDs be offset such that check-node computations are in
the form $f=1-|\delta|=1-\tanh|\lambda/2|$, so that highly certain messages are near $f=0$.
Since the check-node update for LD (\ref{LD_CNupdate}) is odd in every input and $\sign(\delta)=\sign(\lambda)$,
we may simply handle the sign at the conclusion of the check-node-message calculation.

To begin the derivation of the check-node computation for the offset-likelihood difference (OLD) algorithm we note that
\begin{equation*}
f_{l}^{[i \leftarrow j]} = 1- \prod_{k \in \mathcal{N}(j)\setminus i} \left| \delta_{l - 1}^{[k \to j]}  \right|
\end{equation*}
simplifies significantly when performed pairwise.
Letting $\mathcal{N}(j)\setminus i = \{k_1,k_2\}$, the check-node computation becomes
\begin{equation}
\label{OLD_CNup2}
f_{l}^{[i \leftarrow j]} = f_{l - 1}^{[k_1 \to j]}+f_{l - 1}^{[k_2 \to j]}-f_{l - 1}^{[k_1 \to j]}\cdot f_{l - 1}^{[k_2 \to j]}.
\end{equation}

If we wish to perform the variable-node update in the LLR domain, then we note that the transformations
between domains are relatively simple, since
\begin{equation}
\label{OLD_CNup_fromLLR}
f_i = \frac{2{e^{ -|\lambda_i|}}}{1+{e^{ -|\lambda_i|}}}= \frac{2}{1+{e^{|\lambda_i|}}}
\end{equation}
and
\begin{equation}
\label{OLD_CNup_toLLR}
|\lambda_i| = \ln\left(\frac{2-f_i}{f_i}\right).
\end{equation}

\begin{algorithm}
\caption{Offset-LD Check-Node Update for One Edge}
\label{algoff}
\begin{algorithmic}[1]
\Procedure{offset\_ld\_cn}{$n,\vect{\lambda}$} 
\State $f\gets 0$  \Comment{$f$ holds the magnitude, $0<f\le1$}
\State $s\gets +1$  \Comment{$s$ holds the sign, $s\in\{+1,-1\}$}
\For{$i\gets 1, n$}
	\State $g\gets 2* { \exp(-|\lambda[i]|)}/ \left[{1+\exp(-|\lambda[i]|)}\right]$
	\State $f\gets f+g-f*g$
	\State $s\gets s*\sign(\lambda[i])$
\EndFor
\State $\lambda_{out}\gets s * \ln\left(\frac{2-f}{f}\right)$
\State \textbf{return} $\lambda_{out}$
\EndProcedure
\end{algorithmic}
\end{algorithm}

The pairwise calculation of (\ref{OLD_CNup2}) generalizes easily to the recursion of Algorithm \ref{algoff}.
By definition, the range of $f$ is the interval $(0,1]$.
For simplicity Algorithm \ref{algoff} is written to compute a single output message; however, a check node must compute an output for each edge.
Thus, the transformed LLR values from (\ref{OLD_CNup_fromLLR}) on line 5 of Algorithm \ref{algoff} may be pre-computed (just once) for all output messages.
Furthermore, we may organize the pairwise computations on a trellis \cite{HuSPA}.

The computational complexity of Algorithm \ref{algoff} is less than that of (\ref{CNupdate2}); however the dynamic range is also less.
The numerical dynamic range is limited due to \emph{underflowing} (\textit{i.e.}, rounding to zero) $2e^{-|\lambda|}$ in (\ref{OLD_CNup_fromLLR}).
This was shown in Section \ref{sectGIT2} to occur as LLRs exceed $(emax+p)\ln 2$.

The other numerical concern is overflowing $2/f$ in (\ref{OLD_CNup_toLLR}) when $f$ is so small that it is significantly denormalized.
However, for small $f$ (\textit{e.g.}, $f_i <2^{-p}$), we may accurately restate  (\ref{OLD_CNup_toLLR}) as $|\lambda_{i}| = \ln2 -\ln f_i$, which 
has no major numerical concerns and fewer operations.

\subsection{Summary of SPA Formulations}
Table \ref{table_summ} summarizes the findings on LLR limits with respect to check-node inputs.
The traditional formulations based on $\tanh$ operations in (\ref{CNupdate}), Gallager's involution transform, and LD are all very limited in their usable LLR range.
Our preferred approach  (\ref{CNupdate2}) has a range 306 orders of magnitude greater.  

\begin{table}[!t]
\renewcommand{\arraystretch}{1.15} 
\caption{Upper LLR limits of SPA Formulations with respect to Check Node Inputs}
\label{table_summ}
\centering
\begin{tabular}{l||l||r||r}
\hline
\bfseries Tech- & \bfseries LLR-equivalent  & \bfseries LLR limit  & \bfseries LLR limit  \\
\bfseries  nique    & \bfseries limit (approx.) & \bfseries for DP-FP & \bfseries for QP-FP \\
\hline\hline
(\ref{CNupdate})     & $(p+2) \ln 2$ & $38.12$ & $79.72$ \\  
(\ref{CNupdate2})   & $2^{emax+1}$ & $1.798 \times 10^{308}$ & $1.190 \times 10^{4932}$\\
MSA    & $2^{emax+1}$ & $1.798 \times 10^{308}\rlap{\textsuperscript{a}}$ & $1.190 \times 10^{4932}\rlap{\textsuperscript{a}}$\\
GIT     & $(p+2) \ln 2$ & $38.12$ & $79.72$ \\
GIT2   & $(emax+p) \ln 2$ & $745.8$ & $11434$ \\
LR       & ${(emax+1)\ln 2}/{2}$ & $354.9$ & $5678$ \\
LD       & ${(p+1)}\ln2$ & $37.43$ & $79.02$ \\
OLD    & $(emax+p) \ln 2$ & $745.8$ & $11434$ \\
\hline
\multicolumn{4}{l}{\scriptsize \textsuperscript{a}MSA decoder approximates SPA with a performance loss of $0.6$ to $1.22$ dB.}
\end{tabular}
\end{table}

Fig. \ref{fig_cnnoise} shows the output accuracy of a CN using $64$-bit FP calculations for several SPA formulations.
The MSA approximation peaks at low LLRs and then improves substantially, 
while offset-BP is more accurate than MSA at low LLR and less accurate at high LLR.
The approximation by Richter {et al.} \cite{Richter05} has substantially less error than the other approximations throughout the LLR range.
Finally, the vertical lines indicate two exact formulations reaching their upper LLR limit and their error beginning a linear growth with respect to LLR.

\section{Hybrid SPAs and rescaling}
\begin{figure}
\centering
\includegraphics[width=\figwidth]{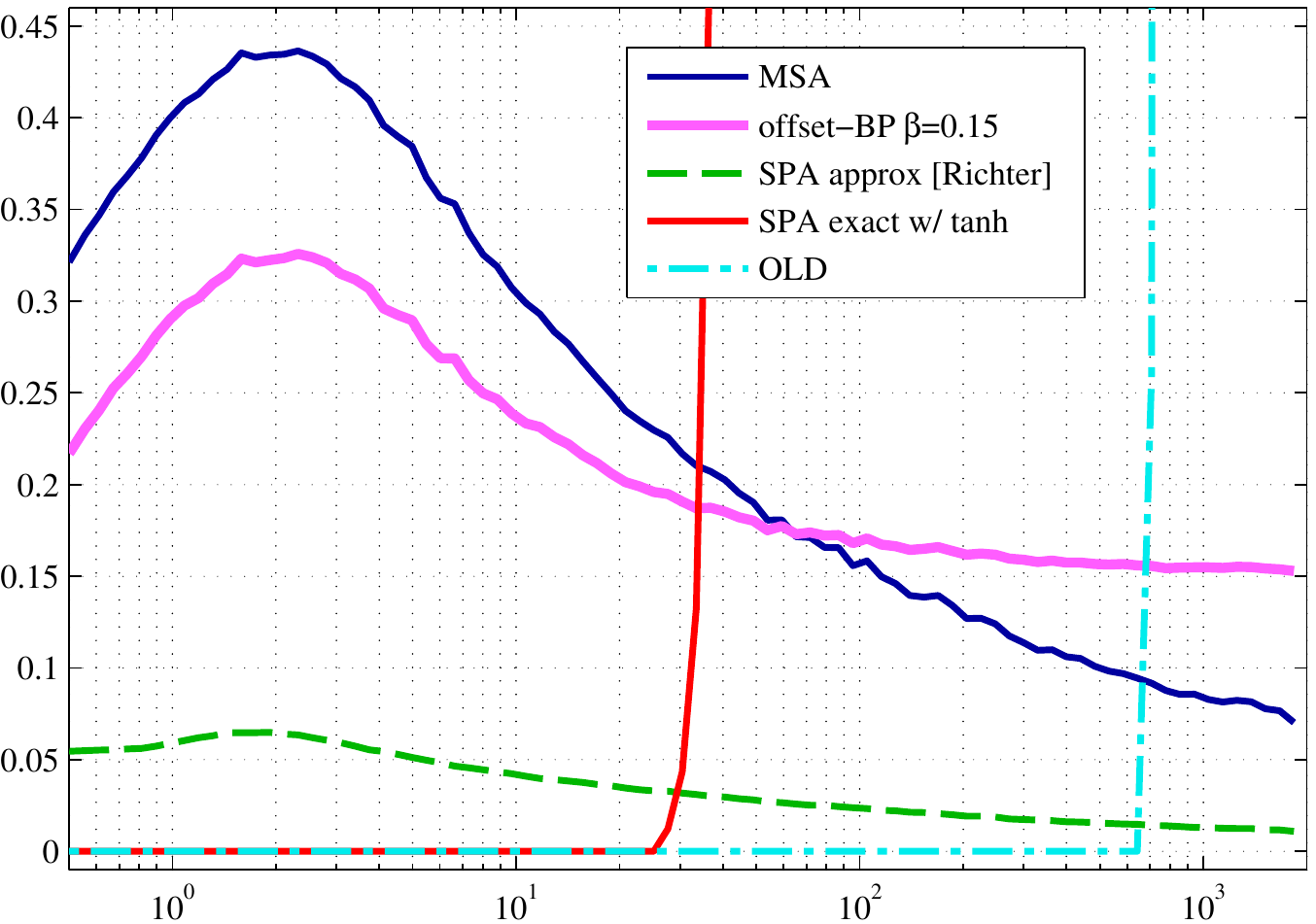}
\caption{RMS LLR error versus mean input LLR value, $m_\lambda$, for check node calculation 
with $4$ input LLRs having i.i.d. Gaussian distribution with variance $\sigma^2_\lambda=2m_\lambda$.}
\label{fig_cnnoise}
\end{figure}

Fig. \ref{fig_cnnoise} also suggests that hybrid SPA solutions -- that is, switching from a precise formulation to an approximation as LLRs grow --  may be acceptable.
In fact, one of our SPA decoder implementations in DP-FP switches from the Richter {et al.} approximation to MSA once a very large LLR, say $2^{p+3}$, is first seen.
At such a point in decoding, the approximation error of the less-complex MSA is less than the resolution of the values for most of the messages in the graph.

Also, it is simple to push the LLR limits beyond that shown in Table \ref{table_summ} for formulations that already support large LLRs.
At very large LLRs the update rules are largely insensitive to scaling, and MSA is always insensitive to scaling.
So, once we detect an extremely large LLR in decoding, say $10^{305}$ in DP-FP, we may rescale all LLRs (including the LLRs from the channel) 
substantially to provide additional headroom.  
This alleviates the need to consider the QP-FP format for more LLR range.
We recognize that, when quantizing FP in the LLR domain (as we prefer) and rescaling,  the quantization steps 
get larger as LLR magnitudes increase.  
We believe this effect to be tolerable when operating in the error floor region as LLRs grow exponentially in trapping set conditions \cite{SunPhD,ButlerAller,ButlerFloor}.

\begin{figure}
\centering
\includegraphics[width=\figwidth]{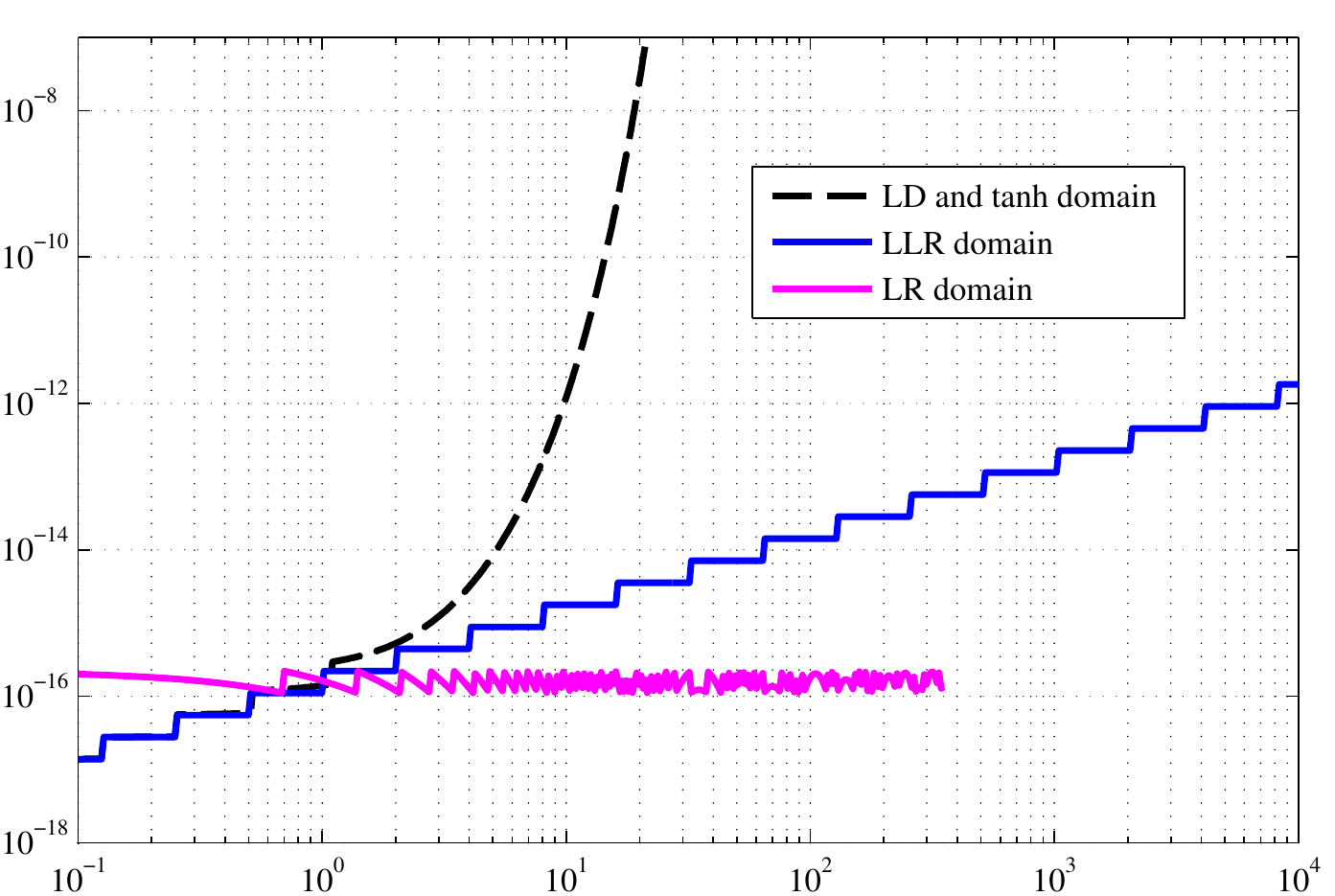}
\caption{$64$-bit FP resolution in LLR units versus equivalent LLR $\lambda$.}
\label{fig_cnresol}
\end{figure}

\section{Floating-Point Resolution}
\label{sectFPResol}
Since range is not the only issue in quantization, we briefly examine floating-point (FP) resolution for the several domains covered.
Fig.~\ref{fig_cnresol} shows the DP-FP resolution (or quantization step-size) of several domains plotted in terms of LLR resolution versus
equivalent LLR.  
Since the curve for ``LLR domain'' is in its native domain, the resolution takes on discrete values.
It is approximately proportional to the value represented, with a proportionality constant of $\alpha=2^{-52.5}=1.6\times10^{-16}$.

We can plot the other domains using a simple transformation.
For instance, in the LR domain ($L=e^\lambda$) of Section \ref{sectLR}, we find the FP-quantization steps of $L$ as a function of $\lambda$ are approximately
\begin{equation*}
\Delta L \approx \alpha L = \alpha e^\lambda \text{ in LR units}.
\end{equation*}
To transform small changes in $L$ to LLR units we need only multiply by the magnitude of the 
derivative $|d \lambda / {d L}|=1/L$ to produce 
\begin{equation*}
\left|\frac{d \lambda}{d L}\right| \Delta L \approx \alpha L / L = \alpha \text{ in LLR units},
\end{equation*}
approximately a constant value.
As shown in Fig.~\ref{fig_cnresol} the true LR resolution in LLR units dithers about our approximate result
until it runs out of range.
The GIT2 and OLD domains yield a resolution overlapping the LR result, 
but the step-size of GIT2 is smaller at LLRs less than $0.2$. 
For the LD or $\tanh$ domain we perform the same analysis and produce the top curve in Fig.~\ref{fig_cnresol}, which shows the step-size growing rapidly.  

\section{Simulation}
\begin{figure}
\centering
\includegraphics[width=\figwidth]{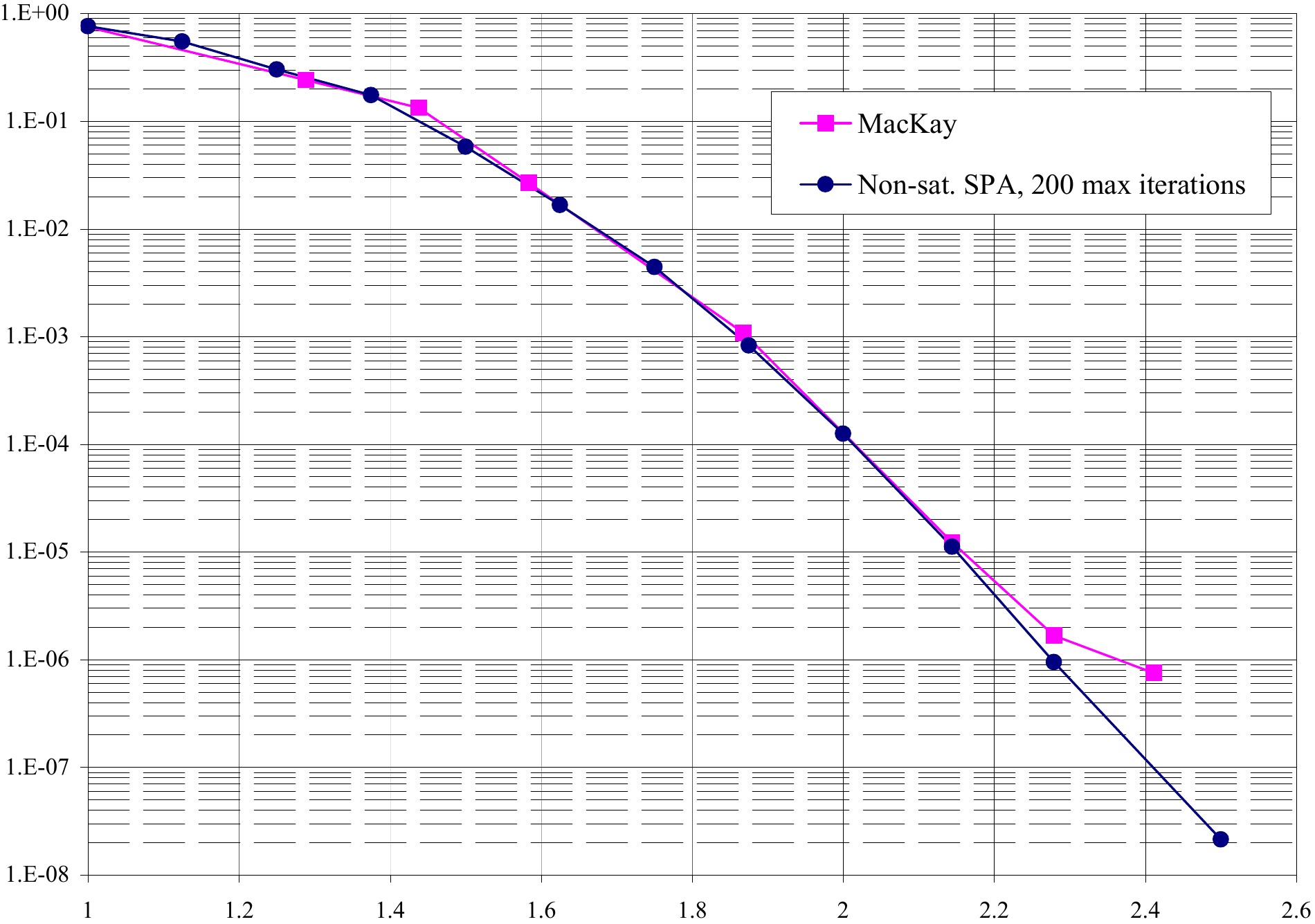}
\caption{FER vs. $E_b/N_0$ in dB for the Margulis LDPC code in AWGN.}
\label{fig_Marg}
\end{figure}

We present performance simulation results for two codes to demonstrate the techniques of this paper.
Fig.~\ref{fig_Marg} shows the frame error rate (FER) of the $(2640,1320)$ Margulis code, 
which is a $(3,6)$-regular LDPC code.  
We show MacKay and Postol's \cite{MKfloor} results in addition to our own.
While they show the start of an error floor at $10^{-6}$ at $E_b/N_0 = 2.4$ dB, other
studies have found the floor to be substantially higher \cite{RyanLin,RichFloors,Han09}.
MacKay and Postol note that this error floor is caused by certain near codewords \cite{MKfloor}.
Our own simulation, using a non-saturating SPA decoder running for a maximum of 200 iterations,
showed no sign of a floor down to $10^{-8}$ and significantly lower.
In fact, resorting to techniques similar to importance sampling we found an error-rate contribution
due to the supposedly dominant near codewords of just $2\times 10^{-11}$ at $E_b/N_0 = 2.8$ dB for this decoder \cite{ButlerFloor}.

\begin{figure}
\centering
\includegraphics[width=\figwidth]{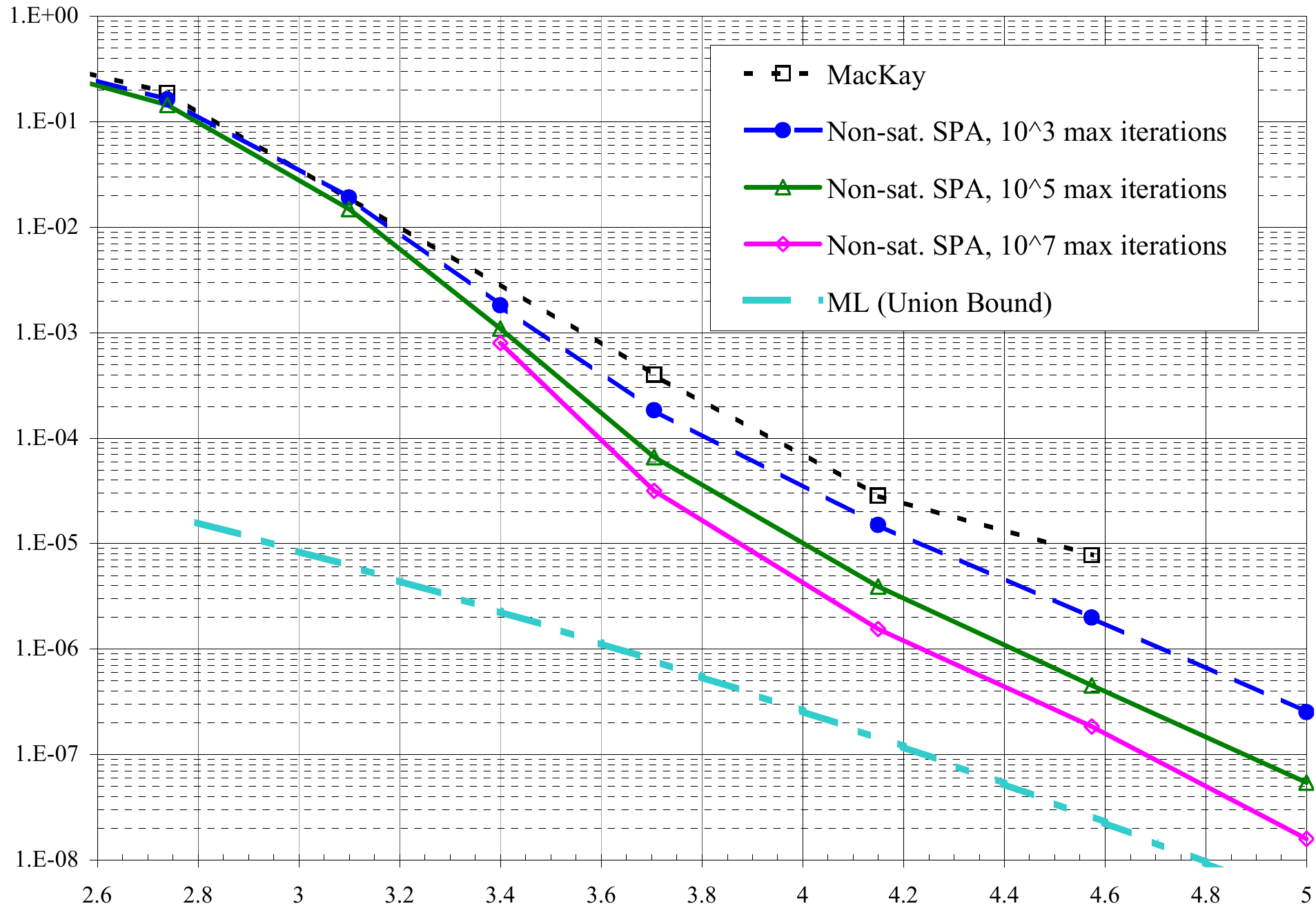}
\caption{FER vs. $E_b/N_0$ in dB for $n=1057$ LDPC code in AWGN.}
\label{fig_DJCM}
\end{figure}

Fig.~\ref{fig_DJCM} shows the FER results for an LDPC code listed as 1057.244.3.457 in \cite{MKency}. 
It is a $(3,13)$-regular code, with block length $n=1057$, rate $R\approx0.77$, and $d_{min}=8$.
Again, we compare our results to those of MacKay. We used our non-saturating SPA decoder with three settings for the maximum number of iterations: 
$10^3$, $10^5$, and $10^7$.  
The average number of decoder iterations performed at $E_b/N_0 = 4.574$ dB was $2.290$, $2.344$, and $4.08$, respectively.
Despite the small increase in average iterations required, significant performance improvements were obtained by using the larger
maximum number of iterations.

\section{Conclusion}
We have addressed numerical limitations on the allowable size of log-likelihood ratios (LLRs) in floating-point (FP) simulations of several formulations of the sum-product algorithm (SPA).
We have described preferred techniques to accommodate very large LLR values and even unbounded range through rescaling.
Additionally, we have proposed simple numerical improvements to Gallager's involution transform that extend its dynamic range by a factor of about $20$ (for double-precision FP computations).
Finally, we introduced a new exact SPA formulation, ``offset-likelihood difference,'' which supports a moderate LLR range with low computational complexity.

\section*{Acknowledgment}  
The authors would like to thank Kenneth Andrews and Ido Tal for helpful discussions.



\bibliographystyle{IEEEtran}



\end{document}